\documentclass[
    ,final            
  ]
  {aipproc}

\layoutstyle{8x11single}

\def\eq#1{{Eq.~(\ref{#1})}}

\def\g{\sqrt{-g}\,}

\newcommand{\LL}{Lanczos-Lovelock}

\newcommand{\Cal}[1]{\ensuremath{\mathcal{#1}}}
\newcommand{\ph}[1]{\phantom{#1}}
\newcommand{\D}{\ensuremath{\nabla}}
\newcommand{\AltC}[8]{\ensuremath{\delta^{#1 #2 #3... #4}_{#5 #6 #7
      ... #8}}} 
\newcommand{\Alt}[6]{\ensuremath{\delta^{#1 #2 ... #3}_{#4 #5
      ... #6}}} 
       
\newcommand{\Riem}[4]{\ensuremath{R^{#1 #2}_{#3 #4}}}
\newcommand{\LDm}{\ensuremath{L_{(m)}}}
\newcommand{\sD}[1]{\sum_{m=1}^{K}{#1}}
\newcommand{\dV}{\ensuremath{\partial\Cal{V}}}
\newcommand{\cc}{cosmological constant}

\usepackage{color}
\usepackage{amssymb}

\begin{document}

\title{A PHYSICAL INTERPRETATION OF GRAVITATIONAL FIELD EQUATIONS}


\author{T. Padmanabhan}{
  address={IUCAA, Pune University Campus,\\
  Ganeshkhind, Pune 411007, INDIA.\\
  email: nabhan@iucaa.ernet.in
  }
}

\begin{abstract}
 It is possible to provide a thermodynamic interpretation for  the field equations in any diffeomorphism invariant theory 
 of gravity. This insight, in turn,  leads us to the possibility of deriving
 the gravitational field equations from another variational principle without using the metric as a dynamical variable. I review this approach and discuss its implications.
\end{abstract}


\maketitle


\section{1. Motivation and Summary}

An unsatisfactory feature of all theories of gravity is that the field
equations do not have any direct physical interpretation. 
The lack of an elegant principle which can lead to the dynamics of gravity
(``how matter tells spacetime to curve'')
is quite striking when we compare this situation with the kinematics of gravity (``how spacetime makes the matter move''). The latter can be determined through the principle of equivalence by demanding that all freely falling observers, at all events in spacetime, must find that the equations of motion for matter  reduce to their special relativistic form. 
Our first aim will be to remedy this and provide a physical interpretation to 
field equations describing gravity in any diffeomorphism invariant theory. 
This, in turn, will lead us to the possibility of \textit{deriving} the gravitational
field equations from a thermodynamic variational principle without using the metric as a dynamical variable.

The alternative interpretation is based on the thermodynamics of horizons.
Several recent investigations have shown that there is indeed a deep connection between gravitational dynamics and horizon thermodynamics (for a review, see \cite{reviews}). For example, studies have shown that: 
\begin{itemize}
\item 
Gravitational field equations in \textit{a wide variety of theories}, when evaluated on a horizon, reduce to  a thermodynamic identity $TdS=dE+PdV$. This result, first pointed out in ref.\cite{tpsstds}, has now been demonstrated \cite{tds} in several cases 
like the  stationary
axisymmetric horizons and evolving spherically symmetric horizons
in Einstein gravity, static spherically symmetric
horizons and dynamical apparent horizons in
Lovelock gravity, and three dimensional BTZ black hole
horizons, FRW cosmological
models in various gravity
theories and even \cite{hwg} in the case Horava-Lifshitz Gravity. It is not possible to understand, in the conventional approach, why the field equations should encode information about horizon thermodynamics.
\item

Gravitational action functionals in a wide class of theories have a a surface term and a bulk term. In the conventional  approach, we \textit{ignore} the surface term completely
(or cancel it with a counter-term) and obtain the field equation
from the bulk term in the action. Any solution to the field equation obtained 
by this procedure is logically independent of the nature of the surface term.
But  when the \textit{surface term} (which was ignored) is evaluated at the horizon that arises
in any given solution, it  gives the entropy of the horizon! (Again, this result extends far beyond Einstein's theory to situations in which the entropy is not just proportional to horizon area.)
This is possible only because there is a specific holographic relationship \cite{holo,ayan}
between the 
surface term and the bulk term which, however, is  an unexplained feature in the conventional
approach to gravitational dynamics.
Since the surface term has the thermodynamic interpretation as the entropy of horizons,
and is related holographically to the bulk term, we are again led to 
an indirect connection between spacetime dynamics and horizon thermodynamics.
\end{itemize}

Based on these features --- \textit{which have no explanation in the conventional approach} --- one can argue  that there  is a  conceptual reason why we need to relate horizon thermodynamics with gravitational dynamics (in a wide class of theories far more general than just Einstein gravity) and revise  our perspective towards spacetime.
To set the stage for this future discussion,  we begin by recalling the implications of the existence of 
temperature for horizons.

In the study of normal macroscopic systems --- like, for example, a solid or a gas --- one can \textit{deduce} the existence of microstructure just from the fact that the object can be heated.  This was  the insight of Boltzmann which led him to suggest that heat is essentially a form of motion of the microscopic constituents of matter. That is, the existence of temperature  is sufficient for us to infer the existence of microstructure even without any direct experimental evidence. 

The non-zero temperature of a horizon  shows that we can actually heat up a spacetime, just as one can heat up a solid or a gas. An unorthodox way of doing this would be to take some amount of matter and arrange it to collapse and form a black hole. The Hawking radiation
 \cite{Hawking:1975sw} emitted by the black hole can be used to heat up, say, a pan of water just as though the pan was kept inside a microwave oven.
In fact the same result can be achieved by just accelerating through the inertial vacuum carrying the pan of water which will eventually be heated to a temperature proportional to the acceleration  \cite{Davies:1975th}. These processes show that the temperatures of all horizons are as real as any other temperature  \cite{Padmanabhan:2002ha}. Since they give rise to  a class of hot  spacetimes, 
it follows \textit{\`{a} la} Boltzmann that 
the spacetimes should possess microstructure.

In the case of a solid or gas, we know the nature of this microstructure from atomic and molecular physics. Hence, in principle, we can work out the thermodynamics of these systems from the underlying statistical mechanics. This is not possible in the case of spacetime because we have no clue about its microstructure. However, one of the remarkable features of thermodynamics
--- in contrast to statistical mechanics ---
is that the thermodynamic description is fairly insensitive to the details of the microstructure and can be developed as a fairly broad frame work. For example, a thermodynamic identity like $TdS = dE+PdV$ has a universal validity and the information about a \textit{given} system is only encoded in the form of the entropy functional $S(E,V)$. In the case of normal materials, this
entropy arises because of our coarse graining over microscopic degrees of freedom which are not tracked in the dynamical evolution. In the case of spacetime, the existence of horizons for a particular class of observers makes it mandatory that these observers  integrate out degrees of freedom hidden by the horizon. 

 To make this notion clearer, let us start from the principle of equivalence which allows one to construct local inertial frames (LIF) with coordinates $X^i$, around any event in an arbitrary curved spacetime.
  Given the LIF, we can now construct a local Rindler frame (LRF)
  by  boosting along one of the directions  with an acceleration $\kappa$, thereby
   locally transforming the metric to a form given by:
\begin{equation}
 ds^2 = -dT^2 + dX^2 + d\mathbf{x}_\perp^2=-\kappa^2 x^2 dt^2 + dx^2 + dL\mathbf{x}_\perp^2
= - 2 \kappa l \ dt^2 + \frac{dl^2}{2\kappa l}  + d\mathbf{x}_\perp^2
\label{surfrind}
\end{equation} 
where $T=x \sinh (\kappa t);  X= x \cosh (\kappa t)$ and $l=(1/2)\kappa x^2$.
The observers at rest (with $x=$ constant) in the LRF will perceive the $X=T$ null surface as a horizon $\mathcal{H}$ (fig 1).
 These local Rindler observers and the freely falling inertial observers will attribute
   different thermodynamical properties to matter in the spacetime. For example, they will attribute different temperatures and entropies to the vacuum state
   as well as excited states of matter fields.
  When some matter with energy $\delta E$ moves close to the horizon --- say, within a few Planck lengths because it formally takes infinite Rindler time for matter to actually cross $\mathcal{H}$ --- the local Rindler observer will consider it to have  transfered an entropy  $\delta S = (2\pi/\kappa) \delta E$ 
   to the horizon degrees of freedom. We will  show (in Sec. 3) that, when the metric satisfies the field equations of any diffeomorphism invariant theory, this transfer of entropy can be given  \cite{tp09papers} a  geometrical interpretation as the change in the  entropy of the horizon.

This result allows us to associate an entropy functional with the null surfaces which the local Rindler observers perceive as  horizons. 
We can now demand that the sum of the horizon entropy and the entropy of matter that flows across the horizons (both as perceived by the local Rindler observers), should be an extremum for all observers in the spacetime.
This leads to a constraint on the geometry of spacetime which can be stated, in $D=4$, as 
\begin{equation}
 (G_{ab} - 8\pi T_{ab}) n^a n^b =0
 \label{Eenn}
\end{equation} 
for all null vectors $n^a$ in the spacetime.
The general solution to \eq{Eenn} is given by $G_{ab} = 8\pi T_{ab} + \rho_0 g_{ab}$
where $\rho_0$ has to be a constant because of the conditions $\nabla_a G^{ab} =0 = \nabla_aT^{ab}$.
Hence the thermodynamic principle leads uniquely to Einstein's equation with a cosmological constant in 4-dimensions. Notice, however, that \eq{Eenn} has a new symmetry which 
the standard Einstein's theory does not posses; viz., it is invariant under the transformation $T_{ab} \to T_{ab} + \lambda g_{ab}$. This has important implications for the cosmological constant problem which we will discuss in Sec. 4.3.
In $D>4$, the same entropy maximization 
 leads to a more general class of theories called \LL\ models (see Sec. 4.2).

The constraint on the background geometry in \eq{Eenn} arises from our demand that the thermodynamic extremum principle should hold for \textit{all} local Rindler observers.   
This is identical  to the manner in which  freely falling observers are used to determine how gravitational field influences matter. Demanding the validity of special relativistic laws for the 
 matter variables, as determined by all the freely falling observers, allows us to 
determine the influence of gravity on matter. In a similar manner, demanding the maximization of entropy of horizons (plus matter), as measured by all local Rindler observers, leads to the dynamical equations of gravity.

In this approach, both the entropy of horizons as well as the entropy of matter flowing across the horizon will be observer dependent thereby introducing a new level of observer dependence in thermodynamics. In particular, observers in different states of motion will have different regions of spacetime accessible to them; for example, an observer falling into a black hole will not perceive its horizon in the same manner as an observer who is orbiting around it. Therefore we are forced to accept that the notion of entropy is an observer dependent concept. (At a conceptual level this is no different from the fact different freely falling observers will measure physical quantities differently; but in this case, standard rules of special relativity allow us to translate the results between the observers. We do not yet have  a similar set of rules for quantum field theory in noninertial frames.)
All these features  suggest a deep relationship between quantum theory, thermodynamics and gravity which forms the main theme of this article. 

The rest of the article is organized as follows: 
In the next section, I briefly review some of the background results needed for our 
discussion. Section 3 describes an interpretation of gravitational field equations
in a general, diffeomorphism invariant, theory of gravity. Using these results, it 
is possible to introduce an entropy maximization principle from which one can obtain
the same field equations without using the metric as a dynamical variable.
This is done in Sec. 4 and the last section summarizes the results.

\section{2. Review of some standard results} 

We begin by summarizing some of the standard results we will need later on. In particular, we will briefly review: (i)  the field equations in a general class of theories of gravity as well as (ii)  the relation between Noether charge and gravitational entropy of horizons.

\subsection{2.1 A general class of theories of gravity}
 
Consider a theory for gravity described by the metric $g_{ab}$ coupled to matter with some degrees of freedom generally denoted by $q_A$. We will take the action describing such a theory in $D-$dimensions to be
\begin{equation}
A=\int d^Dx \sqrt{-g}\left[L(R^{ab}_{cd}, g^{ab})+L_{matt}(g^{ab},q_A)\right]
\label{genAct}
\end{equation}  
where $ L$ is some scalar built from metric and curvature  and $L_{matt}$ is the matter Lagrangian depending on the metric and  matter variables $q_A$. (We have assumed that   $L$ does not involve derivatives of curvature tensor, to simplify the discussion.) Varying $g^{ab}$ in this action with suitable boundary conditions, we will  get the  equations of motion (see e.g. Refs. \cite{ayan,mohut}):
\begin{equation}
2E_{ab} -   T_{ab}=0
\label{eabminustab}
\end{equation}
where
\begin{equation}
E_{ab}=P_a^{\phantom{a} cde} R_{bcde} - 2 \nabla^c \nabla^d P_{acdb} - \frac{1}{2} L g_{ab};\quad
P^{abcd} \equiv \frac{\partial L}{\partial R_{abcd}}
\label{genEab}
\end{equation}

When $\nabla_a P^{abcd} =0$ these field equations  describe a class of theories, called the \LL\ theories which have very interesting geometrical features \cite{lovelock}.  In this case, we have
\begin{equation}
E_{ab}=P_a^{\phantom{a} cde} R_{bcde}  - \frac{1}{2} L g_{ab};\quad
P^{abcd} \equiv \frac{\partial L}{\partial R_{abcd}}
\label{genEab1}
\end{equation}
The crucial difference between \eq{genEab} and  \eq{genEab1} 
is  that, the $E_{ab}$ in
\eq{genEab1} contains no derivatives of the metric higher than second order thereby leading to field equations which are second order in the metric. In contrast, \eq{genEab} can contain up to fourth order  derivatives of the metric.

The Lagrangians which lead to 
the constraint $\nabla_a (\partial L/\partial R_{abcd})=0$ and hence to 
 \eq{genEab1}
are, of course, quite special. They can be written  as a sum of terms, each involving products of  curvature tensors with the $m-$th term being a product of $m$ curvature tensors. The 
general \LL\ Lagrangian has the form,
\begin{equation}
{L} = \sD{c_m\LDm}\,~;~{L}_{(m)} = \frac{1}{16\pi}
2^{-m} \Alt{a_1}{a_2}{a_{2m}}{b_1}{b_2}{b_{2m}}
\Riem{b_1}{b_2}{a_1}{a_2} \cdots \Riem{b_{2m-1}}{b_{2m}}{a_{2m-1}}{a_{2m}}
\,,  
\label{twotwo}
\end{equation}
where the $c_m$ are arbitrary constants and \LDm\ is the $m$-th
order \LL\ Lagrangian.
The $m=1$ term is proportional to $\delta^{ab}_{cd}R^{cd}_{ab} \propto R$ and leads
to Einstein's theory. 
It is conventional to take $c_1 =1$ so that
the ${\ensuremath{{L}_{(1)}}}$,  reduces to $R/16\pi$.
The normalizations for $m>1$ are somewhat arbitrary for individual \LDm\ since the $c_m$s 
are unspecified at this stage.
The $m=2$ term gives rise to what is known as Gauss-Bonnet theory.
Because of the determinant tensor, it is obvious that in any given dimension $D$ we can only have $K$ terms where 
$2K\leq D$.
It follows that, if $D=4$, then only the $m=1, 2$ are non-zero.
Of these, the Gauss-Bonnet term (corresponding to $m=2$) gives, on variation of the
action, a vanishing bulk contribution in $D=4$.
(In dimensions $D=5 $ to  8, one can have both the Einstein-Hilbert term
and the Gauss-Bonnet term and so on.)

\subsection{2.2 Noether charge and the horizon entropy}

Since our  aim is to 
 provide a thermodynamic interpretation of the field equations in \eq{eabminustab},  we first need an expression for horizon entropy  in this  theory. This has been already obtained by Wald \cite{Wald:1993nt} but we will introduce it in a manner appropriate for our purpose.

In any generally covariant theory, the infinitesimal coordinate transformations $x^a \to x^a + \xi^a$ lead
to conservation of a Noether current that can be obtained as follows:
The variation of the gravitational Lagrangian
resulting from arbitrary variations of $\delta g^{ab}$ generically has to a surface term and hence can be expressed in the
form,
\begin{equation}
\delta(L\sqrt{-g }) =\sqrt{-g }\left( E_{ab} \delta g^{ab} + \nabla_{a}\delta v^a\right). \label{variationL} 
\end{equation}
When the variations in $\delta g^{ab}$  arise through the diffeomorphism $x^a \rightarrow x^a + \xi^a$ we have, $\delta (L\sqrt{-g} ) = -\sqrt{-g} \nabla_a (L \xi^a)$, with $\delta g^{ab} = (\nabla^a \xi^b + \nabla^b \xi^a)$. Substituting these in \eq{variationL} and using the (generalized) Bianchi identity $
\nabla_a E^{ab} = 0 
$, we obtain the
conservation law $\nabla_a J^a = 0$, for the current,
\begin{equation}
J^a \equiv \left(L\xi^a + \delta_{\xi}v^a + 2E^{ab} \xi_b \right) 
\label{current}
\end{equation}
where $\delta_{\xi}v^a$ represents the boundary term which arises for the specific variation
of the metric in the form $ \delta g^{ab} = ( \nabla^a \xi^b + \nabla^b \xi^a$). 
It is also convenient to introduce the antisymmetric tensor $J^{ab}$ by $J^a = \nabla_b J^{ab}$.
Using the known expression for $\delta_{\xi}v^a $ in \eq{current}, it is possible to write an explicit expression for the current $J^a$ for any diffeomorphism invariant theory. For the general
class of theories we are considering, the $J^{ab}$ and $J^a$ can be expressed \cite{mohut} in the form
\begin{equation}
J^{ab} = 2 P^{abcd} \nabla_c \xi_d - 4 \xi_d \left(\nabla_c P^{abcd}\right)
\label{noedef}
\end{equation} 
\begin{equation}
J^a = -2 \nabla_b \left (P^{adbc} + P^{acbd} \right ) \nabla_c \xi_d + 2 P^{abcd} \nabla_b \nabla_c \xi_d - 4 \xi_d \nabla_b \nabla_c P^{abcd} 
\end{equation} 
where $P_{abcd}\equiv (\partial L/\partial R^{abcd})$. These expressions simplify significantly at any event $\mathcal{P}$ where  $\xi^a$ behaves like an (approximate) Killing vector and satisfies the conditions
\begin{equation}
 \nabla_{( a} \xi_{b)} = 0;\quad  \nabla_a \nabla_b \xi_c = R_{c b a d} \xi^d
 \label{cond1}
\end{equation}
(which a true Killing vector will satisfy everywhere).
Then one can easily prove that
$
\delta_{\xi}v^a=0
$
at the event $\mathcal{P}$;  
the expression for Noether current simplifies considerably and is given by
\begin{equation}
 J^a \equiv \left(L\xi^a + 2E^{ab} \xi_b \right). 
 \label{current1}
\end{equation}

By considering physical processes involving horizons, it can be shown that the first law of black hole dynamics, for example, is consistent with the identification of the following expression as the horizon entropy:
\begin{equation}
S_{\rm Noether} =  \beta \int d^{D-1}\Sigma_{a} J^{a}= \frac{1}{2}\beta \int d^{D-2}\Sigma_{ab} J^{ab}
\label{noetherint}
\end{equation} 
where $\beta^{-1}=2\pi/\kappa$ is the temperature of the horizon.
 In the final expression in \eq{noetherint} the integral is over any surface with $(D-2)$ dimension which is a spacelike cross-section of the Killing horizon on which the norm of $\xi^a$ vanishes.
 As an example, consider  the special case of  Einstein gravity for which \eq{noedef} reduces to
\begin{equation}
J^{ab} = \frac{1}{16\pi} \left( \nabla^a \xi^b - \nabla^b \xi^a\right)
\end{equation}
If $\xi^a$ be  the timelike Killing vector in the  spacetime describing  a Schwarzschild black hole, we can compute the Noether charge $Q$  as an integral of $J^{ab}$  over any
two surface  which is a spacelike cross-section of the Killing horizon on which 
the norm of $\xi^a$ vanishes. The area element on the horizon can be taken to be  $d\Sigma_{ab} = (l_a\xi_b-l_b\xi_a)\sqrt{\sigma}d^{D-2}x$ with $l_a$ being an auxiliary vector field satisfying the condition $l_a\xi^a=-1$. Then the integral in \eq{noetherint} reduces to
\begin{equation}
 S_{Noether}= -\beta\frac{1}{8\pi}\int\sqrt{\sigma}d^{D-2}x
 (l_a\xi_b)\nabla^b \xi^a
 =\beta\frac{1}{8\pi}\kappa\int\sqrt{\sigma}d^{D-2}x
 =\frac{1}{4}A_H
\end{equation} 
where $A_H$ is the horizon area. (We have used  the relations $ \xi^a \nabla_a \xi^b = \kappa \xi^b, \
l_a\xi^a=-1,\ \beta\kappa=2\pi$ and the fact that $\xi^a$ is a Killing vector.) This result, of course, agrees with the standard one.

The expression for entropy in \eq{noetherint} allows one to interpret $\beta_{loc} J^a$, where $\beta_{loc}$
is the appropriately redshifted local temperature near the horizon, as an entropy density associated with the horizon. This is the interpretation which we will exploit in what follows.

\section{3. The meaning of gravitational field equations}

With this background, we are now in a position to provide 
 a thermodynamic interpretation for
 the gravitational field equations  in the local rindler frames (LRFs) around any event. We will do this in a manner analogous to the way we use the  freely falling observers to determine the kinematics of gravity. At every event in spacetime, we will introduce local Rindler observers and use the horizon thermodynamics
perceived by these Rindler observers to constrain the background geometry.
We shall begin by making the notion of local Rindler observers and their coordinate systems well defined. 

\subsection{3.1 Local Rindler frame and its horizon}
 
Let us choose any event $\mathcal{P}$ and introduce a local inertial frame (LIF) around it with Riemann normal coordinates $X^a=(T,\mathbf {X})$ such that $\mathcal{P}$ has the coordinates $X^a=0$ in the LIF.  
Let $k^a$  be
 a future directed null vector at $\mathcal{P}$ and we align the coordinates of LIF
 such that $k^a$  lies in the $X-T$ plane at $\mathcal{P}$  (Fig.~1). We next  transform from the LIF to a local Rindler frame (LRF) coordinates $x^a$ by accelerating along the X-axis with an acceleration $\kappa$ by the usual transformation. The metric near the origin now reduces to the form in \eq{surfrind} 
where ($t,x, \mathbf {x}_\perp$) 
are the coordinates of LRF.  
 Let $\xi^a$ be the approximate Killing vector corresponding to translation in the Rindler time such
that the vanishing of $\xi^a\xi_a \equiv -N^2$ characterizes the location of the 
local horizon $\mathcal{H}$ in LRF. Usually, we shall do all the computation 
on a timelike surface infinitesimally away from $\mathcal{H}$
with $N=$ constant, usually called a  ``stretched horizon''. 
 Let the timelike unit normal to the stretched horizon
be  $r_a$.

\begin{figure}
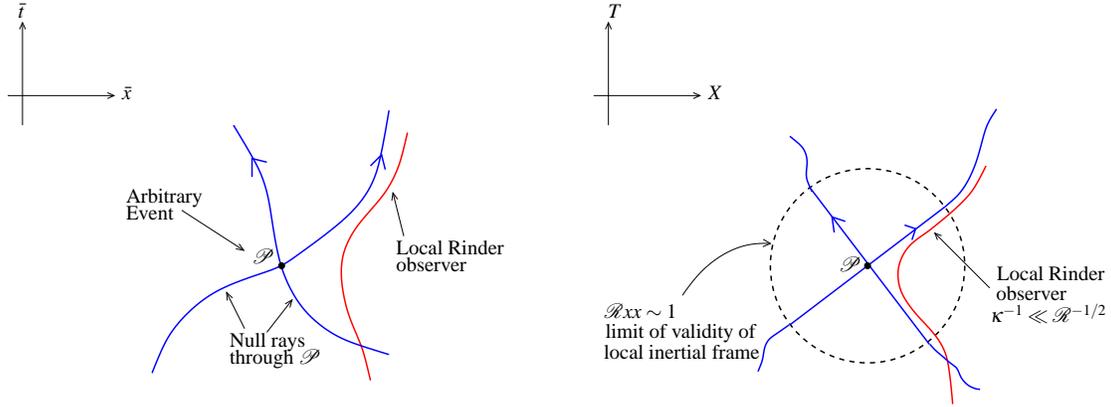

\centering
\scalebox{0.31}{\input{ray1.pstex_t}}\qquad\scalebox{0.31}{\input{ray2.pstex_t}}
\caption{The left frame illustrates schematically the light rays near an event
$\mathcal{P}$ in the $\bar t - \bar x$ plane of an arbitrary spacetime. The right frame
shows the same neighbourhood of $\mathcal{P}$ in the locally inertial frame at $\mathcal{P}$
in Riemann normal coordinates $(T,X)$. The light rays now become 45 degree lines and the trajectory of the local Rindler observer becomes a hyperbola very close to $T=\pm X$ lines
which act as a local horizon to the Rindler observer.}
\end{figure}

\begin{figure}
\scalebox{0.31}{\input{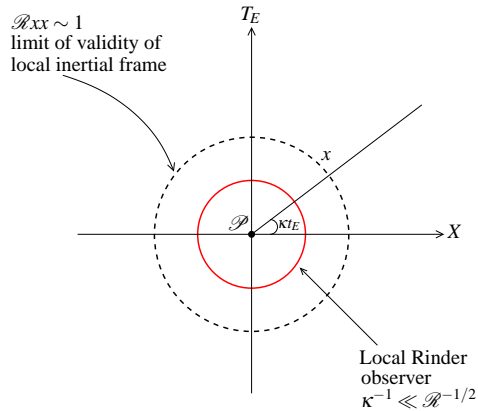}}
\caption{The region around $\mathcal{P} $ shown in figure 1 is represented in the 
Euclidean sector obtained by analytically continuing to imaginary values of $T$ by $T_E = iT$. The horizons $T=\pm X$ collapse to the origin and the hyperbolic trajectory of the Rindler 
observer becomes a circle of radius $\kappa^{-1}$ around the origin. The Rindler coordinates
$(t,x)$ become --- on analytic continuation to $t_E = it$ --- the polar coordinates
$(r=x, \theta = \kappa t_E$) near the origin.}
\end{figure}

 This LRF (with  metric in \eq{surfrind}) and its local horizon $\mathcal{H}$ will exist within a region of size $L\ll\mathcal{R}^{-1/2}$ 
 (where $\mathcal{R}$ is a typical component of curvature tensor of the background spacetime) as long as $\kappa^{-1}\ll\mathcal{R}^{-1/2}$. This condition can always be satisfied by taking a sufficiently large $\kappa$. 
 Essentially,  the introduction of the LRF uses  the fact that we have two length scales in the problem at any event. First is the length scale $\mathcal{R}^{-1/2}$ associated with the curvature components of the background metric over which we have no control. The second is the length scale $\kappa^{-1}$ associated with the accelerated trajectory which we can choose. Hence  we can always ensure that  $\kappa^{-1}\ll\mathcal{R}^{-1/2}$.

 In fact,  this
  is clearly seen in the  Euclidean sector in which the horizon maps to the origin (Fig.~2). The locally flat frame in the Euclidean sector will exist in a region of radius $\mathcal{R}^{-1/2}$ while the 
 trajectory of a uniformly accelerated observer will be a circle of radius $\kappa^{-1}$. Hence one  can always keep the latter inside the former.
 In the Euclidean sector the Rindler observer's trajectory is a circle of radius $\kappa^{-1}$ which can be made arbitrarily close to the origin. Suppose the observer's trajectory has the usual form $X=\kappa^{-1}\cosh \kappa t; T=\kappa^{-1}\sinh \kappa t$ which is maintained for a time interval  of the order of $ 2\pi/\kappa$. Then, the trajectory will complete a full circle \textit{in the Euclidean sector} irrespective of what happens later. When we work in the limit of $\kappa\to\infty$, this becomes arbitrarily local in both space \textit{and} time. 
 The metric in \eq{surfrind} is just the metric of the locally flat region in polar coordinates.

 More generally, one  can choose a trajectory $x^i(\tau)$ such that its acceleration $a^j=u^i\nabla_i u^j$ (where $u^i$ is the time-like four velocity) satisfies the condition
 $a^ja_j =-\kappa^2$. In a suitably chosen LIF this trajectory will reduce to the standard hyperbola of a uniformly accelerated observer. 
 This construction also defines local Rindler horizons around any event.
 Further, the local temperature on the stretched horizon will be $\kappa/2\pi N$ so that 
$\beta_{\rm loc} = \beta N$ with $\beta \equiv 2\pi/\kappa$.

\subsection{3.2 Thermodynamic interpretation of gravitational field equations}
 
The local Rindler observers will view the thermodynamics of matter around them very differently from the freely falling observers (We will say more about this later on in Sec. 3.3). In particular, they will attribute a loss of entropy $\delta S = (2\pi/\kappa) \delta E$  when matter with an amount of energy $\delta E$ gets close to the horizon
 (within a few Planck lengths, say). In the Rindler frame the appropriate energy-momentum  density is $T^a_b\xi^b$. (It is the integral of $T^a_b\xi^b d\Sigma_a$ that gives the Rindler Hamiltonian $H_R$, which leads to evolution in Rindler time $t$ and appears in the thermal density matrix $\rho=\exp-\beta H_R$.) 
 A local Rindler observer, moving along the orbits of the Killing vector field $\xi^a$ with four velocity $u^a = \xi^a/N$, will associate  an energy
  $\delta E =u^a(T_{ab}\xi^b) dV_{\rm prop}$ with a proper volume $dV_{\rm prop}$.
  If this energy gets transfered across the horizon, the corresponding entropy transfer will be $\delta S_{\rm matter} = \beta_{\rm loc}\delta E$ where $\beta_{\rm loc} = \beta N = 
(2\pi/\kappa)N$ is the local  (redshifted) temperature of the horizon
 and $N$ is the lapse function. 
 Since $\beta_{\rm loc} u^a = (\beta N)(\xi^a/N) = \beta \xi^a$, we find that 
\begin{equation}
 \delta S_{\rm matter} = \beta \xi^a \xi^b T_{ab}\ dV_{\rm prop}
 \label{defsmatter}
\end{equation} 
Consider now the gravitational entropy associated with the local horizon. From the 
discussion of Noether charge as horizon entropy  [see \eq{noetherint}], we know that $\beta_{\rm loc} J^a$, 
associated with the (approximate) Killing vector $\xi^a$,
can be thought of 
as local entropy current. Therefore,  $\delta S = \beta_{\rm loc} u_a J^a dV_{\rm prop}$ can be interpreted
as the gravitational entropy associated with a volume $dV_{\rm prop}$ as measured
by an observer with four-velocity $u^a$. (The conservation of  $J^a$ ensures that  there is no irreversible entropy production or dissipation in the spacetime.)  Since $\xi^a$   satisfies \eq{cond1} locally,  it follows that $\delta_\xi v=0$ giving the current to be
$J^a = \left( L \xi^a + 2 E^a_b \xi_b \right) $.
 For observers moving along the orbits of the Killing vector $\xi^a$
with $u^a = \xi^a/N$  we get
\begin{equation}
\delta S_{\rm grav} = \beta N u_a J^a dV_{\rm prop}  =\beta  [\xi^j \xi^a (2E_{aj}) +  L  (\xi_j \xi^j)]\, dV_{\rm prop}
\end{equation} 
As one approaches the horizon, $\xi^a\xi_a\to 0$ making the second term vanish and we find that
\begin{equation}
\delta S_{\rm grav} = \beta  [\xi^j \xi^a (2E_{aj})] \, dV_{\rm prop}
\end{equation}
In the same limit $\xi^j$ will become proportional to the original null vector $k^j$ we started with 
(viz., $\xi^i$ goes to $\kappa \lambda k^i$ where $\lambda $ is the affine parameter associated with the null vector $k^a$ we started with) 
keeping everything finite.   
We now see that the condition $\delta S_{\rm grav} = \delta S_{\rm matter}$ leads
to the result
\begin{equation}
 [2E^{ab} - T^{ab}]k_ak_b =0
 \label{myeqn}
\end{equation} 
Since the original null vector $k_a$ was arbitrary, this equation should hold for all
null vectors for all events in the spacetime.
This is equivalent to $2E^{ab} - T^{ab}=\lambda g^{ab}$ with some constant $\lambda$.
(The constancy of $\lambda$ follows from the conditions $\nabla_aE^{ab}=0,\ \nabla_a T^{ab}=0$.)

 This provides a purely thermodynamical interpretation of the 
 field equations of any diffeomorphism invariant theory of gravity. Note that  
 \eq{myeqn} is not quite the same as the standard equation in \eq{eabminustab} because \eq{myeqn} has an extra symmetry
 which standard gravitational field equations do not have: These equations are invariant
 under the shift $T^{ab}\to T^{ab}+\mu g^{ab}$ with some constant $\mu$. (This symmetry has important implications  \cite{TPgravimmune}  for cosmological constant problem which we will discuss in Sec. 4.3.) While the properties of LRF are relevant conceptually to define the intermediate notions (local Killing vector, horizon temperature ....),  the essential result is independent of these notions. \textit{Just as we introduce local inertial frames to decide how gravity couples to matter, we use local Rindler frames to interpret the physical content of the field equations.}

In the above interpretation we used the local, conserved, current $J^a$. To understand why this is to be expected for consistency, recall that we have crucially used the ``democracy of all observers'' in demanding entropy balance to hold for all observers.
  The mathematical content of  this active version of general covariance is captured by the diffeomorphism
 invariance of the underlying theory which  determines the 
 dynamics of the spacetime.  Because  the  diffeomorphism invariance of the theory forced us to treat all observers at an equal footing,   the diffeomorphism invariance must also lead to   the conserved current $J^a$.  
 We have already seen that this is the case. 
 
The expression for the Noether current is not unique in the sense that one can add to it
any term of the form $Q^a = \nabla_b Q^{ab}$ where $Q^{ab}$ is an anti-symmetric tensor
thereby ensuring $\nabla_a Q^a=0$. This ambiguity has been extensively discussed in the literature but for providing the thermodynamic interpretation to the field equations,
 we have ignored this ambiguity and used the expression in \eq{current}. 
 There are several reasons why the ambiguity is irrelevant for our purpose. First,
 in a truly thermodynamic approach, one specifies the system by specifying a thermodynamic potential, say, the entropy functional. In a local description, this translates into 
 specifying the entropy current which determines the theory. So it is perfectly acceptable to make a specific choice for $J^a$ consistent with the symmetries of the problem. Second, we shall often be interested in theories in which the equations of motion are no higher
 than second order. In these (so called
 \LL\ models) it is not natural to add any extra term to the Noether current such that
 it is linear in $\xi^a$ as we approach the horizon with a coefficient determined entirely from metric and curvature. Finally, we shall obtain in Sec. 4
 the field equation from maximizing an entropy functional where this ambiguity will not 
 arise. 
 
 Once we realize that the real physical meaning of the field equations
 is contained in  \eq{myeqn}, it is possible to re-interpret
 these equations in several alternative ways all of which have the same physical content.
 We shall mention two of them. 
 
 Consider an observer who sees 
 some matter energy flux crossing the  horizon.   Let $r_a$ be the spacelike unit normal to the stretched horizon $\Sigma$, pointing in the direction of increasing $ N$.  If $\xi^a$ is the  approximate, timelike Killing vector corresponding to Rindler time,  then the energy flux through a
 patch of stretched horizon with normal $r_a$ will be $T_{a}^{b}\xi^ar_b$ and the associated entropy flux will be $\beta_{loc} T_{a}^{b}\xi^ar_b$ where $\beta_{loc}^{-1}=\beta^{-1}/N$ is the local temperature.
 We require this entropy flux must match  the entropy change of the locally perceived  horizon.
 The gravitational entropy current is given by  $\beta_{loc}J^a$, such that
 $\beta_{loc}(r_aJ^a)$  gives the corresponding  gravitational entropy flux.  So we require
  $
\beta_{loc} r_a J^a =\beta_{loc} T^a_b r_a \xi^b
$ 
 to hold at 
 all events. 
 The product $r_a J^a$ for the vector $r^a$, which satisfies $\xi^ar_a=0$ on the stretched horizon is
$
r_a J^a = 2 E_{a}^{b}\xi^a r_b  $.
 Hence we get
$
\beta_{loc} r_a J^a =2 E_{a}^{b}  \xi^a r_b =\beta_{loc} T_{a}^{b}\xi^a r_b. 
$
As $ N\to0$ and the stretched horizon approaches the local horizon and $ N r^i$ approaches $\xi^i$ (which in turn is proportional to $k^i$) so that $\beta_{loc}r_a=\beta N r_a\to\beta\xi_a$. So, 
 as we approach the horizon we obtain \eq{myeqn}. 
 
There is another way of interpreting this result which will be useful for further generalizations. Instead of allowing  matter to flow  across the horizon, one could have equally well
  considered a  virtual, infinitesimal (Planck scale), displacement of the $\mathcal{H}$ normal to itself
engulfing some matter. We only need to consider infinitesimal displacements because the entropy of the matter is not `lost' until it crosses the horizon; that is, until when the matter is at an infinitesimal distance (a few Planck lengths) from the horizon. Hence  an infinitesimal displacement of $\mathcal{H}$ normal to itself will engulf 
some matter. 
 Some entropy will be again lost to the  outside observers unless  displacing a piece of local Rindler horizon itself  costs some entropy. 
 We, therefore, expect the entropy balance condition  derived earlier to ensure this and indeed it does.  
An infinitesimal displacement of a local patch of the stretched horizon in the direction of $r_a$, by an infinitesimal proper distance $\epsilon$, will change the proper volume by $dV_{prop}=\epsilon\sqrt{\sigma}d^{D-2}x$ where $\sigma_{ab}$ is the metric in the transverse space.
 The flux of energy through the surface will be  $T^a_b \xi^b r_a$ and the corresponding  entropy flux
 can be obtained by multiplying the energy flux by $\beta_{\rm loc}$.  Hence
 the `loss' of matter entropy to the outside observer because the virtual displacement of the horizon has engulfed some matter is 
$\delta S_m=\beta_{\rm loc}\delta E=\beta_{\rm loc} T^{j}_a\xi^a r_j dV_{prop}$. 
To find the change in the gravitational entropy, we again use the Noether current $J^a$ corresponding
to the local Killing vector $\xi^a$. 
Multiplying by $r^a$ and 
 $\beta_{\rm loc} = \beta N$, we get
\begin{equation}
\beta_{\rm loc} r_a J^a  = \beta_{\rm loc}\xi^a r_b (2E^{b}_a)  + \beta N ( r_a \xi^a) L
\end{equation}  
As the stretched horizon approaches the true horizon,   $N r^a \to \xi^a$
 and $\beta \xi^a \xi_a L \to 0$ making the last term vanish. So the condition $\delta S_{\rm grav} =\delta S_m$ leads to
\begin{equation}
\delta S_{\rm grav} = \beta (2E_{aj})\xi^a \xi^j dV_{prop}
=\delta S_m =\beta T_{aj}\xi^a \xi^j dV_{prop}
\label{localentropy}
\end{equation} 
which is again the same as  \eq{myeqn}. Obviously, there are many other equivalent ways of presenting this result.

\subsection{3.3 Aside: Observer dependence of horizons and entropy}\label{sec:obsdependence}

As an aside, we shall comment on some new conceptual issues brought about by the existence of horizons and entropy which are relevant in this context. We recall that the mathematical formulation leading to the association of temperature with any horizon is fairly universal and it does not distinguish between different horizons, like for example Rindler horizon in flat space or a Schwarzschild black hole event horizon or a de Sitter horizon \cite{Padmanabhan:2002ha}. Assuming that temperature and entropy arise for fundamentally the same reason, it would be extremely unnatural \textit{not} to associate entropy with all horizons.

This feature, however, brings in a  new layer of observer dependent  thermodynamics into the theory which
--- though it  need not come as a surprise ---
has to be tackled. 
We know that while an inertial observer will attribute
  zero temperature and zero entropy to the inertial vacuum, a Rindler observer will attribute a finite temperature and non-zero (formally divergent; `entanglement') entropy to the same vacuum state. \textit{So entropy is indeed an observer dependent concept.} When one does  quantum field theory in curved spacetime, it is 
  not only that particles become an observer dependent notion so do the  temperature and entropy. This notion can be made
  more quantitative as follows: 
  
  Consider an excited state of a quantum field with energy  $\delta E$ above the ground state as specified in an inertial frame. When we integrate out  the unobservable modes for the Rindler observer in this state, we will get a density matrix $\rho_1$ 
  and the corresponding entropy will be $S_1 = - {\rm Tr}\ (\rho_1 \ln \rho_1)$.
  The inertial vacuum state has the density matrix $\rho_0$ and the entropy $S_0 = - {\rm Tr}\ (\rho_0 \ln \rho_0)$. The difference $\delta S = S_1 - S_0$ is finite and 
  represents the entropy attributed to the excited state by the Rindler observer. (This is
  finite though $S_1$ and $S_0$ can be divergent.) In the limit of $\kappa \to \infty$,
  which would correspond to  a Rindler observer who is very close to the horizon,
   we can show that 
  \begin{equation}
\delta S = \beta \delta E = \frac{2\pi}{\kappa} \delta E
\label{delS}
\end{equation}
To prove this, note that if we write $\rho_1=\rho_0 + \delta \rho$, then in the limit of
$\kappa \to \infty$ we can concentrate on states for which $\delta \rho/\rho_0\ll 1$.
Then we have
\begin{eqnarray}
-\delta S &=& {\rm Tr}\ (\rho_1 \ln \rho_1) - {\rm Tr}\ (\rho_0 \ln \rho_0) 
\simeq {\rm Tr}\ (\delta \rho \ln \rho_0)\nonumber\\
&=& {\rm Tr}\ (\delta \rho (-\beta H_R)) = - \beta {\rm Tr}\ \left((\rho_1 -\rho_0)H_R\right) \equiv -\beta \delta E
\end{eqnarray} 
where we have used the facts Tr $\delta \rho \approx 0$ and
$\rho_0 =Z^{-1}\exp(-\beta H_R)$ where $H_R$ is the Hamiltonian for the system in the 
Rindler frame. The last line defines the $\delta E$ in terms of the difference in 
the expectation values of the Hamiltonian in the two states. 
This is the amount of entropy a Rindler observer would claim to be lost when the matter
disappears into the horizon.
(This result can be explicitly proved for, say, one particle excited states of the field
\cite{kkp}.)

One might  have naively thought that the expression for entropy of matter
crossing the horizon 
should consist of its energy $\delta E$ and \textit{its own} temperature $T_{\rm matter}$
rather than the horizon temperature $T_h$. 
But the correct expression is $\delta S = \delta E/T_{h}$;
  the  horizon acts as a system with some internal degrees of freedom and temperature $T_h$ \textit{as far as Rindler observer is concerned} so that when one adds an energy $\delta E$ to it, the entropy change is $\delta S = (\delta E/ T_h)$.
Obviously, a  Rindler
observer (or an observer at rest just outside a black hole horizon) will attribute all  these entropy changes to the horizon she perceives while an inertial observer
(or an observer falling through the Schwarzschild horizon) will see none of these phenomena.
This requires us  to accept the fact that many thermodynamic phenomena needs to be now thought of as specifically observer dependent. For example, if we drop some hot matter into a Schwarzschild black hole, then, when it gets to a few Planck lengths away from the horizon it  would interact with the microscopic horizon degrees of freedom \textit{as far as an outside observer is concerned}. A freely falling observer through the horizon will have a completely different picture. We have learnt to live with this dichotomy as far as elementary kinematics goes; we now  need to do the same as regards thermodynamics and quantum processes. 

As far as an outside observer is concerned, matter takes  an infinite amount of
coordinate time to cross the horizon.
However, this is  irrelevant from a practical point of view.
We have, for example,  considerable evidence of very different nature to suggest Planck length acts as lower bound to the length scales that can be operationally defined and that no measurements can be ultra sharp at Planck scales \cite{tplimitations}. So one cannot really talk about the location of the event horizon ignoring fluctuations of this order. Hence from the operational point of view, we only need to consider matter reaching within few Planck lengths of the horizon to talk about entropy loss which is what we have done
in our discussion.
In fact, physical processes very close to the horizon must play an important role
in order to provide a complete picture of the issues we are discussing. There is 
already some evidence \cite{Padmanabhan:1998jp} that the infinite redshift induced by the horizon plays a crucial
role in this though a mathematically rigorous model is lacking.

\section{4. Gravity: The inside story}\label{sec:gravitystory}
 
 \subsection{4.1 An entropy extremum principle for gravitational field equations}
 
 The  last interpretation of the field equations (see \eq{localentropy})
 given in  Sec. 3.2 is similar to switching from a passive point of view to an active point of view. 
 Instead of allowing matter to fall into the horizon, we made a virtual  displacement of the horizon surface to engulf the matter when it is infinitesimally close to the horizon. But  for the theory to be consistent, this displacement
of the horizon surface degrees of freedom should cost  some entropy. 
 By determining 
the functional form of this entropy density, we should be able to obtain the field equations of gravity through an extremum principle.
Recall that thermodynamics relies entirely on the form of the entropy functional to make predictions.
Hence, if we can determine  the  form of entropy functional for gravity  ($S_{grav}$) in terms of the normals to the null surfaces, then it seems
 natural to demand that  the dynamics should follow from the extremum prescription
$\delta[S_{grav}+S_{matter}]=0$ for \textit{all null surfaces in the spacetime} where $S_{matter}$ is the relevant  matter entropy.

The form of $S_{matter}$ and $S_{\rm grav}$ can be determined as follows.
Let us begin with $S_{matter}$ which  is easy to ascertain from the previous discussion.
If $T_{ab}$ is the matter energy-momentum tensor in a general $D(\ge 4)$ dimensional spacetime then an expression for matter entropy \textit{relevant for our purpose} can be taken to be 
\begin{equation}
S_{\rm matt}=\int_\Cal{V}{d^Dx\sqrt{-g}}
      T_{ab}n^an^b
      \label{Smatt}
\end{equation} 
where $n^a$ is a null vector field.  From our \eq{defsmatter} we see that the entropy density associated with proper 3-volume is $\beta(T_{ab}\xi^a\xi^b)dV_{prop}$ where --- on the horizon --- the vector $\xi^a$ becomes proportional to a null vector $n^a$. 
If we now use the Rindler coordinates in \eq{surfrind} in which $\sqrt{-g}=1$ and 
interpret the factor $\beta$ as arising from an integration of $dt$ in the range $(0,\beta)$ we find that the entropy density associated with a proper four volume is $(T_{ab}n^an^b)$. This suggests treating \eq{Smatt} as the matter entropy. 
 For example, if $T_{ab}$ is due to an ideal fluid at rest in  the LIF then $T_{ab}n^an^b$ will contribute $(\rho+P)$, which --- by Gibbs-Duhem relation --- is just $T_{local}s$ where $s$ is the entropy density and $T_{local}^{-1}=\beta N$ is the properly redshifted temperature with $\beta=2\pi/\kappa$ being the periodicity of the Euclidean time coordinate. Then
 \begin{equation}
 \label{intlim}
\int dS=\int\sqrt{h}d^3x\ s=\int\sqrt{h}d^3x\beta_{\rm loc}(\rho+P)=\int\sqrt{h}Nd^3x \beta (\rho+P)
=\int_0^\beta dt \int d^3x\g T^{ab}n_an_b
\end{equation} 
which matches with \eq{Smatt} in the appropriate limit.

This argument works for any matter source, not necessarily the ones with which we conventionally associate an entropy. What is really relevant is only the \textit{energy} flux close to the horizon from which one can obtain an entropy flux.
We \textit{do} have the notion of \textit{energy} flux across a surface with normal $r^a$ being $T_{ab}\xi^br^a$ which holds for \textit{any} source $T^{ab}$. Given some energy flux $\delta E$ in the Rindler frame, there is an associated entropy flux loss $\delta S=\beta\delta E$ as given by \eq{delS}.  It is \textit{this} entropy which is given by \eq{delS} and \eq{Smatt}. 
The only non-trivial feature in \eq{Smatt} is the integration range for time which is limited to $(0,\beta)$. This is done by considering the integrals in the Euclidean sector and rotating back to the Lorentzian sector but the same result can be obtained working entirely in the Euclidean sector. (There is an ambiguity in the overall scaling of $n^a$ since if $n^a$ is null so is $f(x)n^a$ for all $f(x)$; we will comment on this ambiguity, which turns out to be irrelevant, later on.)

Next, let us consider the expression for $S_{\rm grav}$. We will first describe the simplest possible choice and  then consider a more general expression. The simplest choice is to  postulate $S_{grav}$ to be a quadratic expression \cite{aseementropy} in the derivatives of the normal:  
 \begin{equation}
S_{grav}= - 4\int_\Cal{V}{d^Dx\sqrt{-g}}
    P_{ab}^{\ph{a}\ph{b}cd} \D_cn^a\D_dn^b 
    \label{Sgrav}
\end{equation}  
where the explicit form of $P_{ab}^{\ph{a}\ph{b}cd}$ is ascertained below. Given \eq{Smatt} and \eq{Sgrav} the expression for the total entropy   becomes:
\begin{equation}
S[n^a]=-\int_\Cal{V}{d^Dx\sqrt{-g}}
    \left(4P_{ab}^{\ph{a}\ph{b}cd} \D_cn^a\D_dn^b - 
    T_{ab}n^an^b\right) \,,
\label{ent-func-2}
\end{equation}

The field equations of gravity are to be determined   by extremizing this entropy functional. However, there is one 
 crucial conceptual difference 
between the extremum principle introduced here and the conventional one. Usually, given a set of dynamical variables $n_a$ and a functional $S[n_a]$, the extremum principle will give a set of equations \textit{for} the dynamical variable $n_a$. Here the situation is completely different. We expect the variational principle to hold for   \textit{all} null vectors $n^a$ thereby  leading  to a constraint on  the \textit{background
metric.} Obviously, the functional in \eq{ent-func-2} must be rather special to accomplish this and one needs to impose  restrictions on  $P_{ab}^{\ph{a}\ph{b}cd}$ (and $T_{ab}$, though that condition turns out to be trivial) to achieve this.\index{action principle!thermodynamic approach}

It turns out --- as we shall see below --- that two conditions are sufficient
to ensure this.
First, the tensor $P_{abcd}$ should
have the same algebraic symmetries as the Riemann tensor $R_{abcd}$
of the $D$-dimensional spacetime. 
This condition can be ensured if we define $P_a^{\phantom{a}bcd}$ as
\begin{equation}
P_a^{\phantom{a}bcd} = \frac{\partial L}{\partial R^a_{\phantom{a}bcd}}
\label{condd1}
\end{equation} 
where $L = L(R^a_{\phantom{a}bcd}, g^{ik})$ is some scalar. 
The motivation for this choice
arises from the fact that this approach leads to the same field equations as the one with $L$ as gravitational Lagrangian in the conventional approach (which explains the choice of the symbol $L$).
Second, we will postulate the condition:
\begin{equation}
\D_{a}P^{abcd}=0.
\label{ent-func-1}
\end{equation}
as well as $\D_{a}T^{ab}=0$ which is anyway satisfied by any matter energy-momentum tensor.
One possible motivation for \eq{ent-func-1} is to ensure that the field equations are no higher than second order in the metric.
(If we think of gravity as an emergent phenomenon like elasticity, then $n^a$ is like the displacement field in elasticity. The standard entropy functional \cite{landau7} used in theory of elasticity has the form in \eq{ent-func-1} with coefficients being elastic constants. Here the coefficients are $P^{abcd}$
and the condition in  \eq{ent-func-1}
may be interpreted 
as saying the `elastic constants of spacetime solid' are actually `constants'  \cite{elasticgravity}.) 
This is, however, not a crucial condition and in fact we will see below how this condition in \eq{ent-func-1} can be relaxed.

\subsection{4.2 The field equations}
 
Varying the normal vector field $ n^a$ after adding a
Lagrange multiplier function $\lambda(x)$ for imposing the   condition
$ n_a\delta  n^a=0$, we get 
\begin{equation}
-\delta S = 2\int_\Cal{V} d^Dx\sqrt{-g}
  \left[4P_{ab}^{\ph{a}\ph{b}cd}\D_c n^a\left(\D_d\delta n^b\right)
  - T_{ab} n^a\delta n^b
     - \lambda(x) g_{ab} n^a\delta n^b\right]
 \label{ent-func-3}
\end{equation}
where we have used the symmetries of $P_{ab}^{\ph{a}\ph{b}cd}$ and
$T_{ab}$.  An integration by parts and the
condition $\D_dP_{ab}^{\ph{a}\ph{b}cd}=0$, leads to 
\begin{equation}
-\delta S= 2\int_\Cal{V}{d^Dx\sqrt{-g}\left[-4P_{ab}^{\ph{a}\ph{b}cd}
  \left(\D_d\D_c n^a\right) - ( T_{ab}+ \lambda g_{ab}) n^a\right]\delta n^b}
  +8\int_{\dV}{d^{D-1}x\sqrt{h}\left[k_d
  P_{ab}^{\ph{a}\ph{b}cd}\left(\D_c n^a\right)\right]\delta n^b}
\,,
\label{ent-func-4}
\end{equation}
where $k^a$ is the $D$-vector field normal to the boundary \dV\ and
$h$ is the determinant of the induced metric on \dV.  As usual, in order for
the variational principle to be well defined, we require that the
variation $\delta n^a$ of the  vector field should vanish on the
boundary. The second term in \eq{ent-func-4} therefore vanishes, and
the condition that $S[ n^a]$ be an extremum for arbitrary variations of
$ n^a$ then becomes  
\begin{equation}
2P_{ab}^{\ph{a}\ph{b}cd}\left(\D_c\D_d-\D_d\D_c\right) n^a
-( T_{ab}+\lambda g_{ab}) n^a = 0\,,
\label{ent-func-5}
\end{equation}
where we used the antisymmetry of $P_{ab}^{\ph{a}\ph{b}cd}$ in its
upper two indices to write the first term. The definition of the
Riemann tensor in terms of the commutator of covariant derivatives
reduces the above expression to
\begin{equation}
\left(2P_b^{\ph{b}ijk}R^a_{\ph{a}ijk} -  T{}^a_b+\lambda \delta^a_b\right) n_a=0\,, 
\label{ent-func-6}
\end{equation}
and we see that the equations of motion \emph{do not contain}
derivatives with respect to $n^a$ which is, of course, the crucial point. This peculiar feature arose because
of the symmetry requirements we imposed on the tensor
$P_{ab}^{\ph{a}\ph{b}cd}$. We  need the condition in
\eq{ent-func-6} holds for \emph{arbitrary} null vector fields
$ n^a$. One can easily show\cite{aseementropy} that this requires
\begin{equation}
16\pi\left[ P_{b}^{\ph{b}ijk}R^{a}_{\ph{a}ijk}-\frac{1}{2}\delta^a_b L \right]=
 8\pi T{}_b^a +\Lambda\delta^a_b   
\label{ent-func-71}
\end{equation}
where $\Lambda$ is an arbitrary integration constant. It is also easy to see that 
\eq{ent-func-71} is equivalent to \eq{myeqn} with $E_{ab}$ given by \eq{genEab1}.
One crucial difference between \eq{eabminustab} [along with \eq{genEab1}] and \eq{ent-func-71} is the introduction of the cosmological constant $\Lambda$; we will discuss this  later on.
We mentioned earlier that the expression in \eq{ent-func-2} depends on the overall scaling of $n^a$ which is arbitrary, since $f(x)n^a$ is a null vector if $n^a$ is null. But since the arbitrary variation of $n^a$ (with the constraint $n_an^a=0$) includes scaling variations of the type $\delta n^a=\epsilon(x) n^a$, it is clear that this ambiguity is irrelevant for determining the equations of motion. 

To summarize, we have proved the following. 
 Suppose we start with the Lagrangian in \eq{genAct}, define a $P^{abcd}$ by \eq{condd1} ensuring that it satisfies
\eq{ent-func-1}. 
Varying the metric with this action will lead to  field equations in \eq{genEab1}. We have now proved that we will get the \textit{same} field equations (but with a cosmological constant) if we start with the expression in \eq{ent-func-2}, maximize it with respect to $n^a$ and demand that it holds for all $n^a$.

This result might appear a little mysterious at first sight, but the following alternative description will make clear why this works.
Note that, using the constraints on $P^{abcd}$ we can prove the identity
\begin{eqnarray}
\label{details1}
4P_{ab}^{\ph{a}\ph{b}cd} \D_cn^a\D_dn^b&=&
4\D_c[P_{ab}^{\ph{a}\ph{b}cd} n^a\D_dn^b]-4n^aP_{ab}^{\ph{a}\ph{b}cd} \D_c\D_dn^b
=4\D_c[P_{ab}^{\ph{a}\ph{b}cd} n^a\D_dn^b]-2n^aP_{ab}^{\ph{a}\ph{b}cd} \D_{[c}\D_{d]}n^b\nonumber\\
&=&4\D_c[P_{ab}^{\ph{a}\ph{b}cd} n^a\D_dn^b]-2n^aP_{ab}^{\ph{a}\ph{b}cd} R^b_{\phantom{b}icd}n^i
=4\D_c[P_{ab}^{\ph{a}\ph{b}cd} n^a\D_dn^b]+2n^aE_{ai}n^i
\end{eqnarray} 
where the first equality uses \eq{ent-func-1}, the second equality uses the antisymmetry of $P_{ab}^{\ph{a}\ph{b}cd}$ in c and d, the third equality uses the standard identity for commutator of covariant derivatives and the last one is based on 
\eq{genEab} when $n_an^a=0$ and \eq{ent-func-1} hold. Using this in the expression for
 $S$ in \eq{ent-func-2} and integrating the four-divergence term, we can write
\begin{equation}
S[n^a]=-\int_{\partial\Cal{V}}{d^{D-1}x k_c\sqrt{h}}
(4P_{ab}^{\ph{a}\ph{b}cd} n^a\D_dn^b)
-\int_\Cal{V}{d^Dx\sqrt{-g}}(2E_{ab}-T_{ab})n^an^b
\label{thetrick}
\end{equation}
So, when we vary $S[n^a]$ (ignoring the surface term) we are  effectively varying $(2E_{ab}-T_{ab})n^an^b$ with respect to $n_a$ and demanding that it holds for all $n_a$. 
There is an ambiguity of adding a term of the form $\lambda(x)g_{ab}$ in the integrand of the second term in \eq{thetrick} leading to the final equation
$(2E_{ab}=T_{ab}+\lambda(x)g_{ab})$ but the Bianchi identity $\nabla_aE^{ab}=0$ along with
$\nabla_aT^{ab}=0$ will make $\lambda(x)$ actually a constant. 

It is now clear how we can find  an $S$ for any theory, even if \eq{condd1} does not hold. 
This can be achieved by starting from  the expression  $(2E_{ab}-T_{ab})n^an^b$ as the entropy density,  using \eq{genEab} for $E_{ab}$ and integrating by parts (see the discussion after Eq. (14) in ref.  \cite{tp09papers}). In this case, we get
for $S_{\rm grav}$ the expression:
\begin{equation}
S_{\rm grav} = - 4 \int_V d^Dx\, \sqrt{-g}\, \left[ P^{abcd} \nabla_c n_a \, \nabla_d n_b + (\nabla_d P^{abcd})n_b \nabla_cn_a 
 + (\nabla_c \nabla_dP^{abcd}) n_a n_b\right]
\label{generalS}
\end{equation} 
Varying this with respect to $n^a$ will then lead to the correct equations of motion and --- incidentally --- the same surface term.

While one could indeed work with this more general expression, there are four reasons to prefer the imposition of the condition in  \eq{condd1}.
First,
 it is clear from \eq{genEab} that when $L$ depends on the curvature tensor and
the metric, $E_{ab}$ can depend up to the fourth derivative of the metric if \eq{condd1}
is not satisfied. But when we impose \eq{condd1} then we are led to field equations
which have, at most, second derivatives of the metric tensor which is  a desirable feature.
 Second, 
 as we shall see   below,  with that condition we can actually determine the form of $L$; it turns out that in $D=4$, it uniquely selects Einstein's theory, which  is probably a nice feature. In higher dimensions, it picks out a very geometrical extension of Einstein's theory in the form of \LL\ theories.
Third,
it is difficult to imagine why the terms in \eq{generalS} should occur with very specific coefficients. In fact, it is not clear  why we cannot have derivatives of $R_{abcd}$ in $L$, if the derivatives of $P_{abcd}$ can occur in the expression for entropy. 
Finally, 
if we take the idea of elastic constants being constants, then one is led to \eq{condd1}. 
None of these rigorously exclude the possibility in \eq{generalS} and in fact
this model has been
 explored  recently \cite{sfwu}.
 
So far we have not fixed $P^{abcd}$ so we have not fixed the theory. 
In a
complete theory, the explicit form of $P^{abcd}$ will be determined by the
long wavelength limit of the microscopic theory just as the elastic
constants can --- in principle --- be determined from the microscopic
theory of the lattice. In the absence of such a theory, we need to determine $P^{abcd}$ by general considerations which is possible when $P^{abcd}$ satisfies \eq{condd1}.
Since this condition is identically satisfied by \LL\ models which are known to be unique, our problem can be completely solved by taking the  $P^{abcd}$  as a series in the
 powers of  derivatives of
the metric as:
\begin{equation}
P^{abcd} (g_{ij},R_{ijkl}) = c_1\,\stackrel{(1)}{P}{}^{abcd} (g_{ij}) +
c_2\, \stackrel{(2)}{P}{}^{abcd} (g_{ij},R_{ijkl})  
+ \cdots \,,
\label{derexp}
\end{equation} 
where $c_1, c_2, \cdots$ are coupling constants with the $m$ th order term derived from the \LL\ Lagrangian:
\begin{equation}
\stackrel{(m)}{P}{}_{ab}^{cd}\propto
\AltC{c}{d}{a_3}{a_{2m}}{a}{b}{b_3}{b_{2m}}
\Riem{b_3}{b_4}{a_3}{a_4} \cdots
\Riem{b_{2m-1}}{b_{2m}}{a_{2m-1}}{a_{2m}} 
 =
\frac{\partial\LDm}{\partial R^{ab}_{{cd}}}\,. 
\label{zerothree}
\end{equation}
where $\AltC{c}{d}{a_3}{a_{2m}}{a}{b}{b_3}{b_{2m}}$ is the alternating tensor.
 The lowest order
term depends only on the metric with no derivatives. The next
term depends (in addition to metric) linearly on curvature tensor and the next one will be quadratic in curvature etc.
The lowest order term in \eq{derexp} (which  leads to Einstein's theory) is
\begin{equation}
\stackrel{(1)}{P}{}^{ab}_{cd}=\frac{1}{16\pi}
\frac{1}{2} \delta^{ab}_{cd} =\frac{1}{32\pi}
(\delta^a_c \delta^b_d-\delta^a_d \delta^b_c)
  \,.
\label{pforeh}
\end{equation}
To the lowest order, when we use \eq{pforeh} for $P_{b}^{\ph{b}ijk}$, 
  the \eq{ent-func-71} 
 reduces to Einstein's equations.
 The corresponding gravitational entropy functional \footnote{
 Interestingly, the integrand in $S_{GR}$ has the $Tr(K^2)-(Tr K)^2$ structure. If we think of the $D=4$ spacetime being embedded in a sufficiently large k-dimensional \textit{flat} spacetime we can obtain the same structure using the Gauss-Codazzi equations relating the (zero) curvature of k-dimensional space with the curvature of spacetime.}
is:
 \begin{equation}
S_{\rm GR}[n^a]=\int_\Cal{V}\frac{d^Dx}{8\pi}
   \left(\D_an^b\D_bn^a - (\D_cn^c)^2 \right)
\end{equation} 
 The next order term (which arises from  the  Gauss-Bonnet Lagrangian) in \eq{twotwo} is:
\begin{eqnarray}
\stackrel{(2)}{P}{}^{ab}_{cd}= \frac{1}{16\pi}
\frac{1}{2} \delta^{ab\,a_3a_4}_{cd\,b_3\,b_4}
R^{b_3b_4}_{a_3a_4} 
=\frac{1}{8\pi} \left(R^{ab}_{cd} -
         G^a_c\delta^b_d+ G^b_c \delta^a_d +  R^a_d \delta^b_c -
         R^b_d \delta^a_c\right) 
\label{pingone}
\end{eqnarray} 
and similarly for all the  higher orders terms. None of them can contribute in $D=4$ so we get Einstein's theory as the unique choice if we assume $D=4$. If we assume that $P^{abcd}$ is to be built \textit{only} from the metric, then this choice is unique in all $D$.

  While the matter term in the functional in \eq{ent-func-2}  has a natural interpretation
in terms of entropy transfered across the horizon, the interpretation of the gravitational part needs to be made explicit. The interpretation of $S_{\rm grav}$ as entropy  arises from the following two facts. First, we see from the identity \eq{details1} that this term differs from $2E_{ij} n^i n^j$ by a total divergence. On the other hand, we have seen earlier that the term $2E_{ij} n^i n^j$
can be related to the gravitational entropy of the horizon through the Noether current
which suggests the identification.
In fact, 
 \eq{thetrick} shows that when the equations of motion holds \textit{the total entropy of a bulk region is entirely on its boundary}. Further
if we evaluate this boundary term
\begin{equation}
-S|_{\rm on-shell}=4\int_{\dV}{d^{D-1}x\ k_a\sqrt{h}\,\left(P^{abcd}n_c\D_bn_d\right)}
\label{on-shell-2}
\end{equation} 
(where we have manipulated a few indices using the symmetries of
$P^{abcd}$) 
in the case of a \textit{stationary} horizon which can be locally approximated as Rindler spacetime, one gets exactly the Wald entropy of the horizon \cite{aseementropy}.

\subsection{4.3 Cosmological constant}

The approach outlined above has important implications for the \cc\ problem, which we shall now briefly mention \cite{ccreview}.
In the conventional approach, 
 we start with an action principle which depends on matter degrees of freedom and the metric 
 and vary (i) the matter degrees of freedom to obtain the equations of motion for
matter and (ii) the metric $g^{ab}$ to obtain the field equations of gravity.
The equations of motion for \textit{matter} remain invariant if one adds a constant,
say, $-\rho_0$ to the matter Lagrangian.
However, gravity breaks
this symmetry which the matter sector has and $\rho_0$ appears as a \cc\ term in the field equations of gravity. If we interpret the evidence for dark energy in the 
universe (see ref. \cite{sn}; for a critical look at data, see ref. \cite{tptirthsn1} and references therein)
as due to the cosmological constant, then its value has to be
fine-tuned  to satisfy the observational constraints.
It is not clear why a particular parameter in the low energy  sector has to be fine-tuned in such a manner. 

In the alternative perspective described here,
the functional in \eq{ent-func-2} is clearly invariant under the shift $L_m \to L_m - \rho_0$ or equivalently, $T_{ab} \to T_{ab} + \rho_0 g_{ab}$, 
since it only introduces a term $-\rho_0 n_a n^a =0$ for any null vector $n_a$.
In other words, one \textit{cannot} introduce the cosmological constant
as  a low energy parameter in the action in this approach. We saw, however, 
that the cosmological constant can reappear as an \textit{an integration constant} when the equations 
are solved. The integration constants which appear in a particular solution  have a completely different conceptual status compared to the parameters which appear in the action describing
the theory. It is much less troublesome to choose a fine-tuned value for a particular integration constant in the theory if observations require us to do so.
From this point of view, the cosmological constant problem is considerably less severe
when we view gravity from the alternative perspective. 

This extra symmetry  under the shift 
$T_{ab} \to T_{ab} + \rho_0 g_{ab}$ arises because we are not treating metric  as a 
dynamical variable in an action principle.\footnote{It is sometimes claimed that a spin-2 graviton in the linear limit \textit{has to} couple to $T_{ab}$ in a universal manner, in which case, one will have the graviton coupling to the cosmological constant. In our approach, the linearized  field equations for the spin-2 graviton field $h_{ab}=g_{ab}-\eta_{ab}$, in a suitable gauge, will be $(\square h_{ab} -T_{ab})n^an^b=0$ for all null vectors $n^a$. This equation is still invariant under $T_{ab} \to T_{ab} + \rho_0 g_{ab}$ showing that the graviton does \textit{not} couple to cosmological constant.}
In fact one can state a stronger result \cite{gr06}.
Consider any model of gravity satisfying the following three conditions: (1)
The metric  is varied in a local action to obtain the equations of motion. (2) We demand full general covariance of the equations of motion. (3) The equations of motion for matter sector are invariant under the addition of a constant to the matter Lagrangian.  Then, we can prove a `no-go' theorem that the \cc\ problem cannot be solved in such model. That is, we cannot solve \cc\ problem unless we drop one of these three demands. Of these, we do not want to sacrifice general covariance encoded in (2); neither do we have a handle on low energy matter Lagrangian so we cannot avoid (3). So the only hope we have is to introduce an approach in which gravitational field equations are obtained from varying some degrees of freedom other than $g_{ab}$ in a maximization principle.
This suggests that the so called cosmological constant problem has its roots in our misunderstanding of the nature of gravity.
 
 Our approach  at present is not yet developed far enough to predict the value of the \cc\ .But providing a mechanism in which the \textit{bulk cosmological constant
decouples from gravity} is a major step forward.  It was always thought
that  some unknown symmetry should make the \cc\ (almost) vanish and weak (quantum gravitational) effects which break this symmetry could lead to its small value. \textit{Our approach provides a model which has such symmetry.} The small value of the observed \cc\ has to arise from non-perturbative quantum gravitational effects at the next order, for which we do not yet have a fully satisfactory model. (See, however, Ref. \cite{ccsol}.)

\section{5. Conclusions}

It is useful to distinguish clearly (i) the mathematical results which can be rigorously proved from (ii) interpretational ideas which might evolve when our understanding of these issues deepen.

(i) From a purely algebraic point of view, without bringing in any physical interpretation or motivation, we can prove the following mathematical results: 
\begin{itemize}
\item
Consider a functional of null vector fields $n^a(x)$ in an arbitrary spacetime given by
\eq{ent-func-2} [or, more generally, by \eq{generalS}]. Demanding that this functional is an extremum for all null vectors $n^a$ leads to the field equations for the background geometry given by $(2E_{ab}-T_{ab})n^an^b=0$ where $E_{ab}$ is given by \eq{genEab1} [or, more generally, by \eq{genEab}]. Thus field equations in a wide class of theories of gravity can be obtained from an extremum principle without varying the metric as a dynamical variable.
\item
These field equations are invariant under the transformation 
$T_{ab} \to T_{ab} + \rho_0 g_{ab}$, which 
relates to the freedom of introducing a  \cc\ as an integration constant in the theory. Further, this symmetry forbids the inclusion of a cosmological constant  term in the variational principle by hand as a low energy parameter. \textit{That is, we have found a symmetry which makes the bulk \cc\ decouple from the gravity.} When linearized around flat spacetime, the graviton inherits this symmetry and does not couple to the \cc .
\item
On-shell, the functional in \eq{ent-func-2} [or, more generally, by \eq{generalS}] contributes only on the boundary of the region. When the boundary is a horizon, this terms gives precisely the Wald entropy of the theory.
\end{itemize}
 It is remarkable that one can derive not only Einstein's theory uniquely in $D=4$ but even \LL\ theory in $D>4$ from an extremum principle involving the null normals \textit{without varying $g_{ab}$ in an action functional!}. 
 
(ii) As regards interpretation of these results,  we see from \eq{details1}
that in the case of Einstein's theory, we have a Lagrangian $n^a(\nabla_{[a}\nabla_{b]})n^b$ for a vector field $n^a$
(except for  a surface term)
that becomes vacuous in flat spacetime in which covariant derivatives become partial derivatives. There is  no dynamics in $n^a$ (in the usual sense) but they do play a crucial role. 
This raises the question as to the physical meaning of these null vectors and the interpretation of our approach. As described above, this interpretation is essentially thermodynamical and the physical picture is made of the following qualitative ingredients:
\begin{itemize}
\item
Assume that the spacetime is endowed with certain microscopic degrees of freedom capable of 
exhibiting thermal phenomena. This is just the Boltzmann paradigm: \textit{If one can heat it, it
must have microstructure!}; and one can heat up a spacetime. 
\item
Whenever a class of observers perceive a horizon, they are ``heating up the spacetime''
and the  degrees of freedom close to a horizon  participate  
in a very \textit{observer dependent} thermodynamics. 
Matter which flows close to the horizon
(say, within a few Planck lengths of the horizon) transfers energy to these microscopic, near-horizon, degrees of freedom \textit{as far as the observer who sees the horizon is concerned}. Just as  entropy of a normal system at temperature $T$ 
will change by $\delta E/T$ when we transfer to it an energy $\delta E$, here also an entropy change will occur. (A freely falling observer in the same neighbourhood, of course, will deny all these!)
\item
We proved that when the field equations of gravity hold, one can interpret this entropy change in  a purely geometrical manner involving the Noether current.
From this point of view, the  normals $n^a$ to local patches of null surfaces are related to the (unknown) degrees of freedom that can participate in the thermal phenomena involving the horizon. 
\item
Just as demanding the validity of special relativistic laws with respect to all freely falling observers leads to the kinematics of gravity, demanding the local entropy balance in terms of the thermodynamic variables as perceived by local Rindler observers leads to the field equations of gravity in the form $(2E_{ab}-T_{ab})n^an^b=0$. 
 
\end{itemize}

As stressed in earlier sections, this involves a new layer of observer dependent thermodynamics. At a conceptual level, this may be welcome when we note that 
every key progress in physics involved realizing that something we thought as absolute is  not absolute. With special relativity it was the flow of time and with general relativity it was the concept of global inertial frames and when we brought in quantum fields in curved spacetime it was the notion of particles and temperature.
We now know that the temperature attributed to even vacuum state depends on the observer. It seems necessary to  integrate the entire thermodynamic machinery (involving what we usually consider to be the `real' temperature)
 with this notion of LRFs having their own (observer dependent) temperature. There is scope for further work in this direction.


\begin{theacknowledgments}
I thank Jean-Michel Alimi for organizing a stimulating conference and providing excellent hospitality.
\end{theacknowledgments}

\end{document}